\documentclass{appolb}%
\usepackage{graphicx}
\usepackage{amssymb}
\usepackage{amsmath}
\usepackage{color}%
\usepackage{bm}  
\usepackage{amsfonts}

\begin{document}
\renewcommand*{\thefootnote}{\fnsymbol{footnote}}

\title{Heavy baryons in the chiral quark-soliton model: \\a possibility for exotica?\thanks{Presented 
at 
{\em Excited QCD}, Kopaonik, Serbia, March 10 -- 15, 2018.}}
\author{Michal Praszalowicz
\address{M. Smoluchowski Institute of Physics, Jagiellonian University, \\
ul. S. {\L}ojasiewicza 11, 30-348 Krak{\'o}w, Poland.}}

\maketitle

\begin{abstract}
We discuss possible  interpretation of five excited $\Omega^0_c$ states
 within the Chrial Quark Soliton Model.
 We show that it is
not possible to interpret all five $\Omega^0_c$'s as parity minus excitations 
and argue that
two narrowest states are pentaquarks belonging to
the SU(3) representation $\overline{15}$. 

\end{abstract}


\bigskip
\bigskip


\section{Chiral Quark Soliton Model }

In this report we summarize our recent works
on heavy baryons \cite{Yang:2016qdz}\nocite{Kim:2017jpx,Kim:2017khv}--\cite{Praszalowicz:2018upb}
where we have applied the Chiral Quark Soliton Model ($\chi$QSM)
to the baryonic systems with one heavy quark. An expanded version of this report has been
published in Ref.~\cite{Praszalowicz:2018uoa} where a complete list of reference can be found.
There are two
 other contributions based on   \cite{Yang:2016qdz}\nocite{Kim:2017jpx,Kim:2017khv}--\cite{Praszalowicz:2018upb} 
 that have been
already published elsewhere ~\cite{Praszalowicz:2017cwk,Praszalowicz:2018azt}.

The $\chi$QSM \cite{Diakonov:1987ty} (for  review see Ref.~\cite{Christov:1995vm} and references therein)
 is based on an old argument by Witten \cite{Witten:1979kh} that in the limit of a large number of 
 colors ($N_{\rm  c} \rightarrow \infty$), 
 $N_{\rm val}=N_{\rm  c}$ relativistic valence quarks generate chiral mean fields represented by a distortion of 
 a Dirac sea that in turn interacts with the valence quarks themselves.
The soliton configuration corresponds to the solution of a pertinent Dirac equation for the constituent quarks (with
gluons integrated out) in the mean-field approximation, where the mean fields respect so called {\em hedgehog}
symmetry. This means that
neither spin ($\bm{S}$) nor isospin ($\bm{T}$)  are  {\em good} quantum numbers. Instead a {\em grand spin}  
$\bm{K}=\bm{S}+\bm{T}$ is a {\em good} quantum number. In Ref.~\cite{Yang:2016qdz}, following \cite{Diakonov:2010zz},
we have observed that the same argument holds for $N_{\rm val}=N_{\rm  c}-1$, which allows to replace one 
light valence quark by a heavy quark $Q=c$ or $b$.

\begin{figure}[h]
\centering
\includegraphics[width=9.0cm]{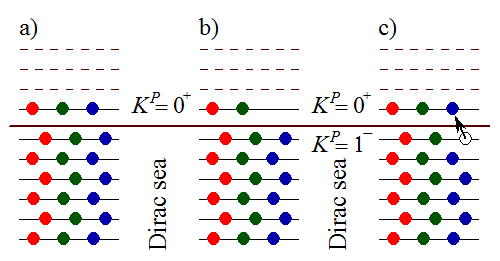} \vspace{-0.2cm}\caption{Schematic
pattern of light  quark levels in a self-consistent soliton
configuration. In the left panel all sea levels are filled and $N_{c}$ (=3 in
the Figure) valence quarks occupy the $K^{P}=0^{+}$ lowest
positive energy level. Unoccupied positive energy levels are dpicted
by dashed lines. In the middle panel one valence quark has been
stripped off, and the soliton has to be 
supplemented by a heavy quark not shown in the Figure. In the right panel a
possible excitation of a sea level quark, conjectured to be $K^{P}=1^{-}$, to
the valence level is shown, and again the soliton has to couple to a heavy
quark. Strange quark levels that exhibit different filling pattern are
not shown.}%
\label{fig:levels}%
\end{figure}

For light baryons the ground state soliton configuration corresponds to the occupied $K^P=0^+$ valence level, 
(with $N_{\rm val}=N_{\rm  c}$) as shown in Fig.~\ref{fig:levels}.a.
Therefore the soliton does not carry definite quantum numbers except for the baryon number resulting from the valence quarks.
It is also possible that one of the valence quarks gets excited to some $K>0$ level (see {\em e.g.} \cite{Petrov:2016vvl}), 
which influences the quantization of the soliton
spin emerging  when the rotations in space and flavor are quantized.
The resulting {\em collective} hamiltonian is analogous to the one
of a symmetric top with the following constraints:
\begin{enumerate}
\item allowed SU(3) representations must contain states with hypercharge
$Y^{\prime}=N_{\rm val}/3$,
\item the isospin $\bm{T}^{\prime}$ of the states with $Y^{\prime}%
=N_{\rm val}/3$ couples with the soliton spin $\bm{J}$ to $\bm{K}$, which is 0 for the ground state
configuration but may be non-zero for an excited state:
$\bm{T}^{\prime}+\bm{J}=\bm{K}$.
\end{enumerate}
For light baryons $N_{\rm val}=N_{c}, \; K^P=0^+$, and
as a result the lowest lying positive parity baryons belong to the SU(3)$_{\rm flavor}$ octet of spin 1/2 and
decuplet of spin 3/2. The first {\rm exotic} representation is  $\overline{\mathbf{10}}$ of spin 1/2 
with the lightest state corresponding to
the putative $\Theta^{+}(1540)$~\cite{Diakonov:1997mm}. The model has been successfully tested in the light 
baryon sector \cite{Christov:1995vm}.

\section{$\chi$QSM and heavy baryons}

\begin{figure}[h]
\centering
\includegraphics[height=5cm]{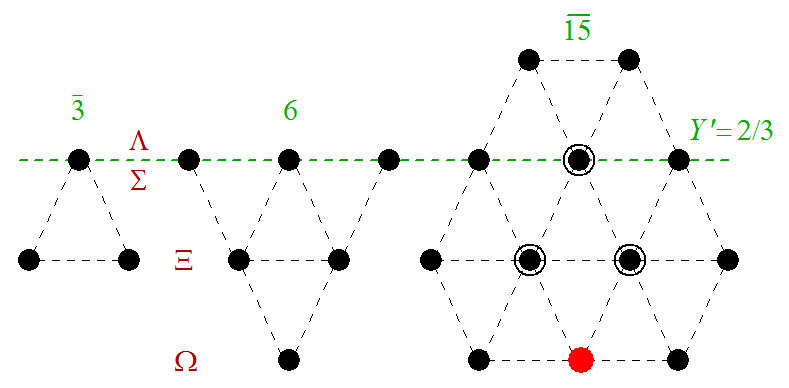}
\caption{Rotational
band of a soliton with one valence quark stripped off. Soliton spin
corresponds to the isospin $T^{\prime}$ of states on the quantization line
$Y^{\prime}=2/3$. We show three lowest allowed representations: antitriplet of
spin 0, sextet of spin 1 and the lowest exotic representation $\overline
{\mathbf{15}}$ of spin 1 or 0.  Heavy quark has to be added.}
\label{fig:irreps}%
\end{figure}

Recently~\cite{Yang:2016qdz} , following Ref.~\cite{Diakonov:2010zz}, we have made a proposal
how to generalize the above approach to heavy baryons,
by stripping off one valence quark from the $K^P=0^+$ level, as shown in Fig.~\ref{fig:levels}.b,
and replacing it by a heavy quark to neutralize the color. 
The only difference to the previous case is the quantization condition, since $N_{\rm val}=N_{\rm  c}-1$.
The lowest allowed SU(3) representations are in this case
 $\overline{\mathbf{3}}$ of spin 0  and 
 ${\mathbf{6}}$ of spin 1 shown in Fig.~\ref{fig:irreps}.

\begin{figure}[h]
\centering
\includegraphics[height=7cm]{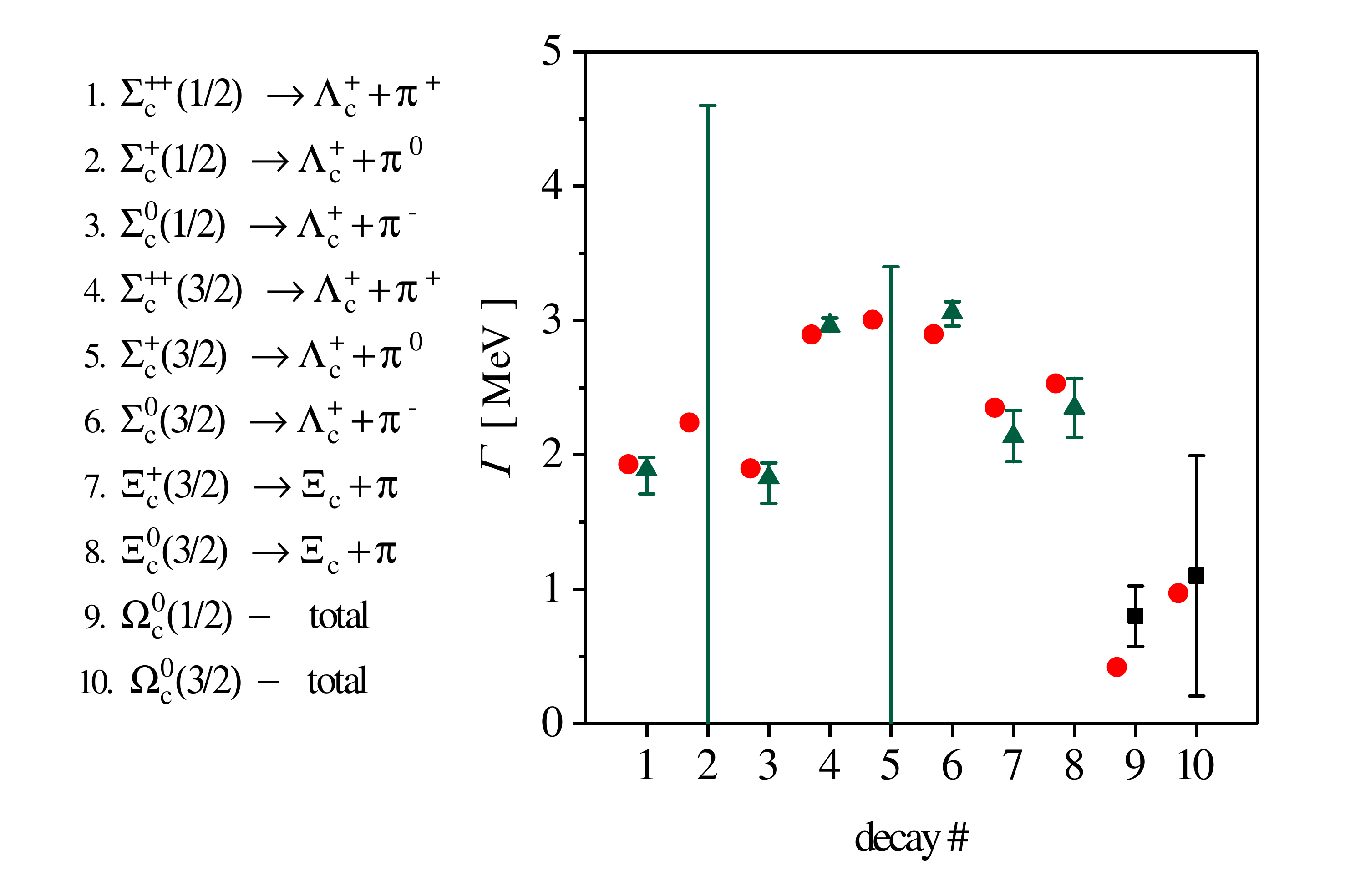}
\caption{Decay
widths of the charm baryons.  Red full circles correspond to our
theoretical predictions. Dark green triangles correspond to the experimental
data~\cite{Patrignani:2016xqp}. 
Data for decays 4 -- 6 of 
$\Sigma_{c}(\mathbf{6}_{1},3/2)$ have been divided by a factor
of 5 to fit within the plot area. Widths of two LHCb 
\cite{Aaij:2017nav}
$\Omega_{c}$ states that we interpret as pentaquarks are plotted as black full
squares with  theoretical values shown as red full circles. }%
\label{fig:charmwidths}%
\end{figure}

As a result
both  ${\mathbf{6}}-\overline{\mathbf{3}}$ splitting and the $m_s$ splittings inside
these multiplets  are {\em predicted} using as an input
the light sector spectrum 
\cite{Yang:2016qdz} except for 
 a hyperfine splitting of {\bf 6} due to the spin-spin interaction of a soliton
and a heavy quark that has been parametrized phenomenologically. Moreover,
we have calculated the decay widths  \cite{Kim:2017khv}, which are in surpriingly
good agreement with the data (see Fig.~\ref{fig:charmwidths} for charm baryons decay widths).

\section{Excitations of heavy baryons}

The $\chi$QSM allows for 
two  kinds of excitations ~\cite{Kim:2017jpx}. Firstly, higher
SU(3) representations, similar to the antidecuplet in the light sector, appear
in the rotational band of the 
soliton of Fig.~\ref{fig:levels}.b.
The lowest possible exotic SU(3) representation is 
$\overline{\mathbf{15}}$ of positive parity  and spin 1
 ($\overline{\mathbf{15}}$ of spin 0 is heavier) 
shown in Fig.~\ref{fig:irreps}. Second possibility corresponds to the
excitation of the sea quark from the $K^P=1^{-}$ sea level to the valence level~\cite{Diakonov:2010zz}
depicted in Fig.~\ref{fig:levels}.b (or alternatively valence quark excitation to the first excited level\footnote{We thank Victor
Petrov for pointing out this possibility.}
of $K^P=1^{-}$). In this case the parity is negative but the rotational band is the same 
as in Fig.~\ref{fig:irreps} with, however, different quantization condition, since $J$ and $T'$ have to couple to $K=1$.

\begin{figure}[h]
\centering
\includegraphics[width=9.0cm]{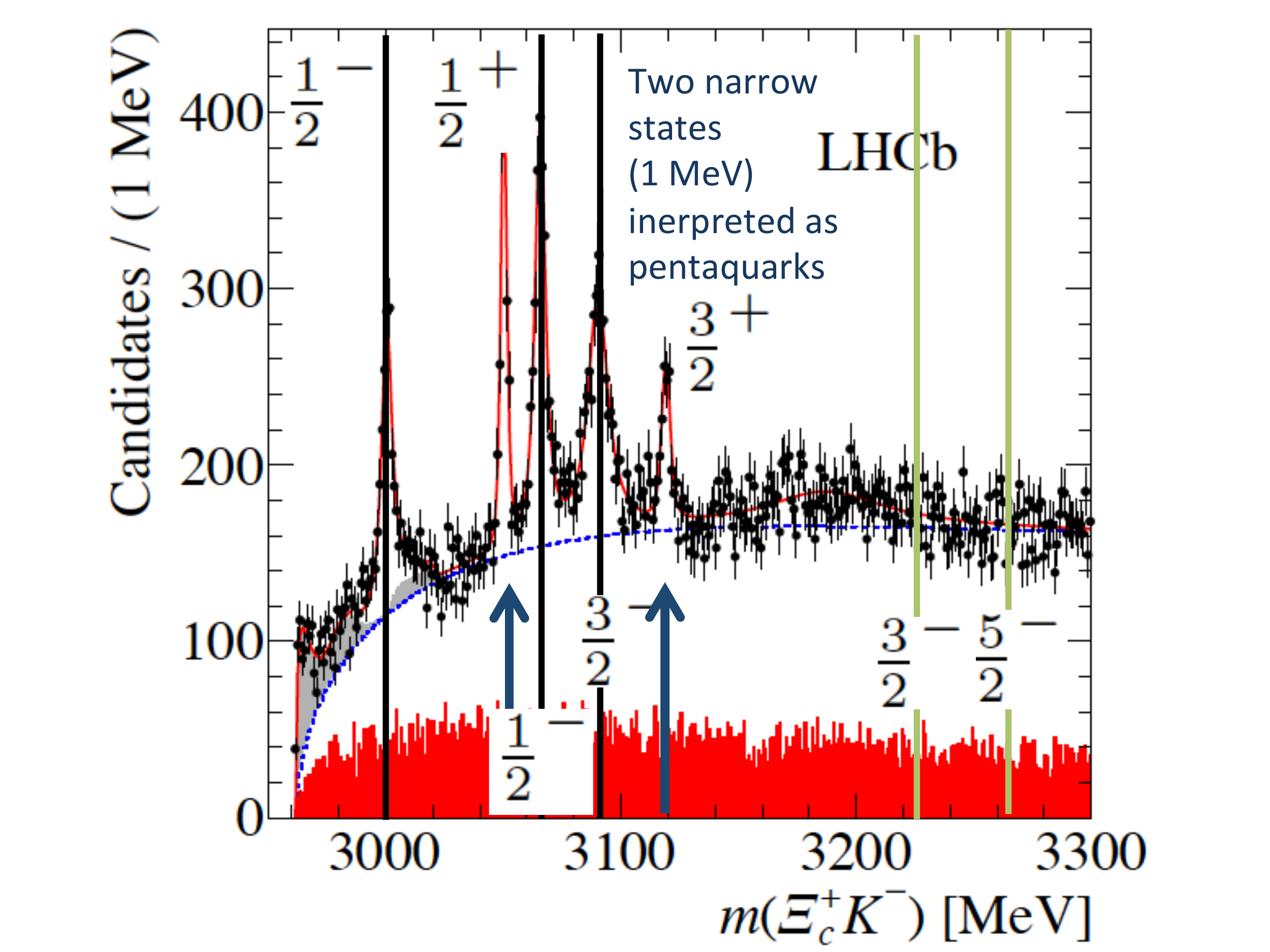} \vspace{-0.2cm}
\caption{Spectrum of the $\Omega^0_c$ states \cite{Aaij:2017nav} with theoretical predictions of the present model}%
\label{fig:Omegas}%
\end{figure}

We have shown that the model describes well the only fully known spectrum of negative parity antitriplets
of spin 1/2 and 3/2~\cite{Kim:2017jpx}. There has been no experimental evidence for the sextet until recent
report of five $\Omega^0_{c}$ states by the LHCb~\cite{Aaij:2017nav} and later by BELLE~\cite{Yelton:2017qxg}.
In the sextet case the quantization condition requires the soliton spin to be quantized
as $J=0,1$ and 2. By adding one heavy quark we end up with five possible total spin $S$ excitations
for $J=0$: $S=1/2$, for $J=1 $: $S=1/2$ and 3/2, 
and for $J=2$:  $S=3/2$ and 5/2. Although the number of states coincides with the experimental 
results~\cite{Aaij:2017nav,Yelton:2017qxg},  it is
not possible to accommodate all five $\Omega^0_c$ states within the constraints imposed by the $\chi$QSM~\cite{Kim:2017jpx}. We have
therefore {\em forced} model constraints (note that in the $\mathbf{6}$ case we cannot predict the mass splittings, since there is a new 
parameter in the splitting hamiltonian that corresponds to the transition of Fig.~\ref{fig:levels}.c, which is not known from the light sector),
which allows to accommodate only three out of five LHCb states (see black vertical lines in Fig.~\ref{fig:Omegas}). 
Two heaviest $\chi$QSM states (green lines in  Fig.~\ref{fig:Omegas}) lie already above the decay threshold to heavy mesons,
and it is quite possible that they have very small branching ratio to the  $\Xi^+_c+K^-$ final state analyzed by the LHCb. Two remaining states
indicated by dark-blue arrows in Fig.~\ref{fig:Omegas}, which are 
hyper fine  split by 70~MeV  (as the ground state sextets that belong to the same rotational band), 
can be therefore interpreted as the members of exotic 
$\overline{\mathbf{15}}$ of positive parity shown as a red dot in Fig.~\ref{fig:irreps}. This interpretation is reinforced
by the decay widths, which can be computed in the model. These widths are of the order of 1~MeV and agree with the
LHCb measurement (see Fig.~\ref{fig:charmwidths}). Such small widths are in fact expected in the present approach,
since the leading $N_{\rm  c}$ terms of the relevant couplings cancel in the non-relativistic limit~\cite{Praszalowicz:2018upb}.

Our identification implies the existence of 
  the {\it isospin} partners of $\Omega^0_c$ in the $\overline{\mathbf
   15}$. They can be searched for in the mass distribution of
 $\Xi_c^0+K^-$ or $\Xi_c^+ + \bar K^0$. Our model applies also to the bottom sector,
 and -- where the data is available -- it describes very well both masses and decay widths~\cite{Yang:2016qdz,Kim:2017khv}.

\section*{Acknowledgments}
It is a pleasure to thank my collaborators H.-C.~Kim, M.V.~Polyakov and G.S.~Yang.
This work was supported by the Polish NCN grant 2017/27/B/ ST2/01314.

\end{document}